\documentclass[twocolumn,showpacs,preprintnumbers,amsmath,amssymb,floatfix]{revtex4}
\usepackage{epsfig}
\usepackage{hyperref}
\usepackage{cleveref}

\usepackage{natbib}
\begin{document}

\title{New insight into EM radiation from spinning dust and its influence on the Cosmic Microwave Background}

\author{Ariel Guerreiro$^{1}$}
\author{J.Tito Mendon\c ca$^{2}$}
\author{Robert Bingham$^{3,4}$}

\affiliation{$^1$Physics Department, Faculdade de Ci\^encias, Universidade do Porto, Rua do Campo Alegre, 687, 4169-007 Porto, Portugal}
\affiliation{$^2$Instituto de Plasmas e Fus\~{a}o Nuclear, Instituto Superior T\'{e}cnico, Av. Rovisco Pais 1, 1049-001 Lisboa, Portugal}
\affiliation{$^3$University of Strathclyde, Glasgow, G4 0NG, UK}
\affiliation{$^4$Rutherford Appleton Laboratory, Chilton, Didcot, Oxfordshire OX11 OQX, UK}

\today

\begin{abstract}

Dust is ubiquitous in the Universe and its influence on the observed Electromagnetic (EM) radiation needs to be correctly addressed. In recent years it became clear that scattering of EM radiation from interstellar dust grains could change the local properties of the observed Cosmic Microwave Background (CMB) radiation. Here we consider the relevant processes of emission and scattering of EM radiation from spinning dust particles, and discuss their possible influence on the CMB. In particular, we show that scattered radiation can establish a correlation between different spectral components of galactic dipolar emission. This could explain the observed correlation between the CMB and the 100-micron thermal emission form interstellar dust.
Another important property of CMB is related with its polarisation anisotropies, and the observation of a cosmological B-mode. We show that scattering of CMB radiation from dust grains in the presence of a static magnetic field could indeed create a B-mode spectral component, which is very similar to that due to primordial gravitational waves. This can be described by a kind of Cotton-Mutton effect on the CMB radiation.

\end{abstract}

\maketitle

\section{Introduction}

It is well known that dust is ubiquitous in the Universe, and their basic properties can be found in \cite{evans1993dusty, whittet2002dust}.
In particular, it was first noticed in 1949 that interstellar dust grains are aligned \cite{hall1949science,hiltner1949science,hiltner1949polarization, hiltner1949presence} and since then, many aspects of dust grain dynamics have been discovered. The influence of dust dynamics on the measurements of the cosmic microwave background (CMB) have mainly been understood as a source of contamination that was necessary to be removed from the observational data. Currently, the analysis of the influence of the dust dynamics on the CMB depend on a large degree of extrapolation and modelling of the rotational dynamics of the grains \cite{resendes2003new}.

Of particular importance in the study of CMB are the anisotropies of their fundamental parameters, most notably the temperature and polarisation anisotropies. In the case of the later, the analysis is done in terms of the B and E modes, measured in terms of two Stokes parameters Q and U that are analogous to decomposing a vector field into a gradient part (E) and a divergence free curl
part (B). The accurate detection of the B and E mode are  extremely important in the understanding of the structure of the early Universe. Cosmological perturbations are typically distinguished between scalar (e.g. energy density perturbations), which produce only E-mode polarisation, or tensor (e.g. gravitational waves), which is evident by the existence of the cosmological B-mode spectral component.

It has been suggested that the B mode signal in BICEP may correspond (totally or partially) to a dust artefact \cite{bucher2015physics}, which reinforces the importance of understanding the impact of interstellar dust on the distribution of the B and E modes in the night sky. More recently, it was noted from satellite observations, that the B-mode signal is compatible with the level of thermal emission from interstellar dust \cite{adam2016planck}, thus making the explanation of the observed B-mode based on dust grain emission more plausible. Here we propose a new mechanism leading to a B-mode signal, which is nearly independent of the dust temperature and only depends on the propagation properties of CMB radiation across magnetized dust grains. 

Much of the dust grain rotational dynamics on its own does not single out any preferred direction and to explain the microwave polarisation requires some mechanism that causes the dust grains to align themselves locally, such as the galactic magnetic field.
The methodology followed here is to derive the equivalent dielectric tensor for each point of space considering that there exist some dust grains with electric and magnetic moments. Then, using special ray-tracing equations, compute the propagation of microwaves through a magnetic pinch generated by a current loop. This allows us to estimate the intensity and polarisation state at some far away image plane.

We consider an extended region of space occupied by a magnetised dust cloud. We assume the presence of a static magnetic field ${\bf B} = (0, 0, B_0)$, such that the dust rotation is constrained to the perpendicular plane. This means that the grain has a magnetic dipole moment $d_q$, with components  ${\bf d}_g = d_q (\cos \phi, \sin \phi, 0)$. This creates an anisotropy in the medium, such that the ordinary and extraordinary modes (polarised parallel and perpendicularly to the static field) will propagate with a different phase velocity. We then consider propagation of depolarised radiation through the medium. The anisotropy will be responsible for partial polarisation of the background radiation, which will increase with the size of the dust cloud, the magnitude of the static magnetic field, and the dipole moment of the dust particles. 

We have preformed ray tracing simulations of initially depolarised radiation and computed the expected changes in the local polarisation states. The observed polarisation outside the cloud, will reflect the eventual changes in magnetic field strength and cloud density. Figure 1 represents our simulation scheme. Details of the propagation model can be found in the Supplementary Material. Simulations of ordinary and extraordinary rays across the dust cloud were performed using some plausible physical assumptions, to be specified later.


\begin{figure}
\centering\includegraphics[width= 1 \columnwidth]{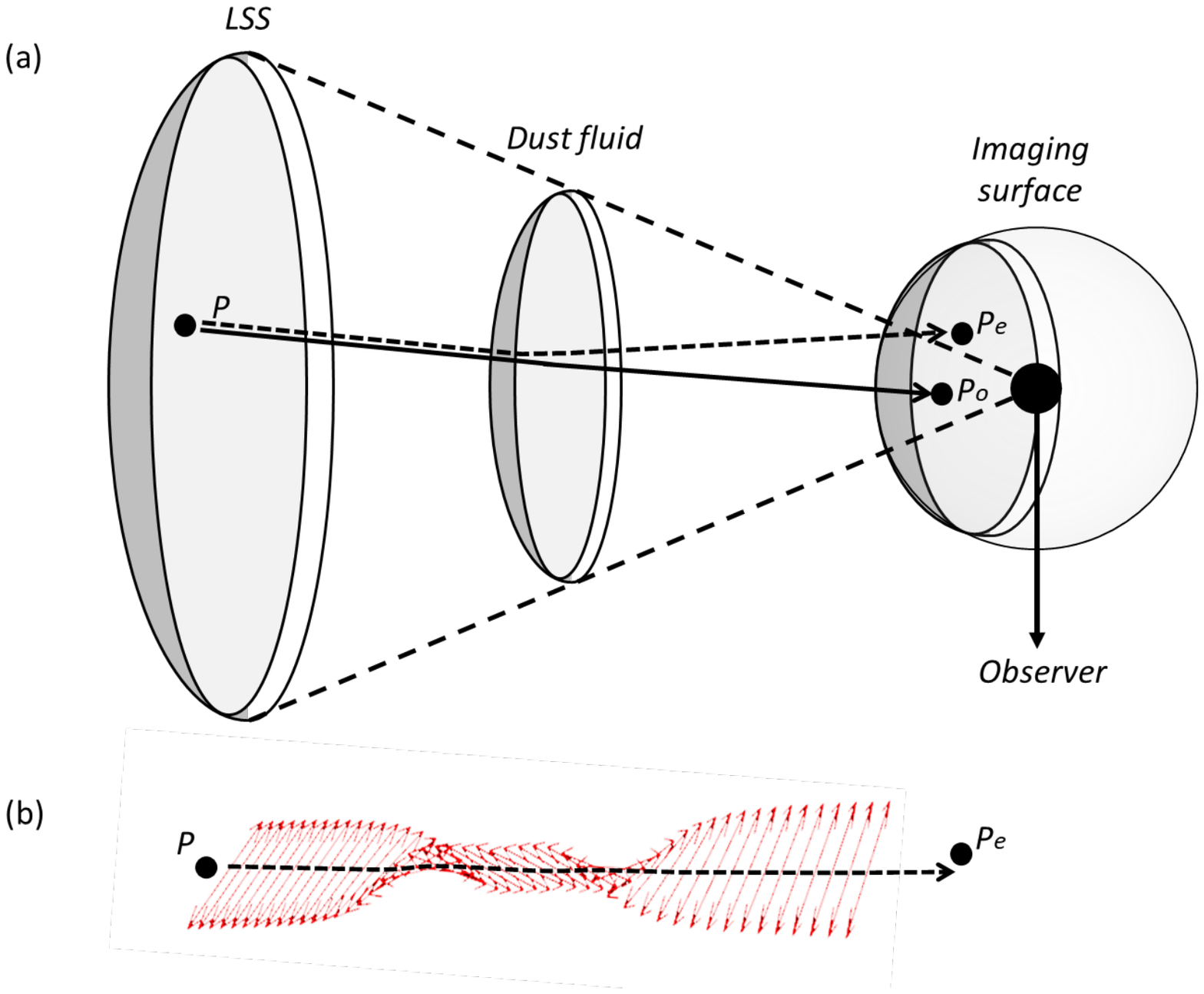}
\caption{Representation of the simulations scheme. (a) A ray is emitted from the last scattering surface (LSS) in the far field at a point P , and propagates through a structure of dust fluid and separated into a ordinary and extraordinary ray, each with a specific polarisation, until they reach the imaging surface near the observer and image at different points, respectively $P_o$ and $P_e$. (b) During propagation of a ray across the dust fluid, the orientation of the polarisation of the ordinary and extraordinary rays may be altered.
}\label{fig1}
\end{figure}

\begin{figure}
\centering\includegraphics[width=1 \columnwidth]{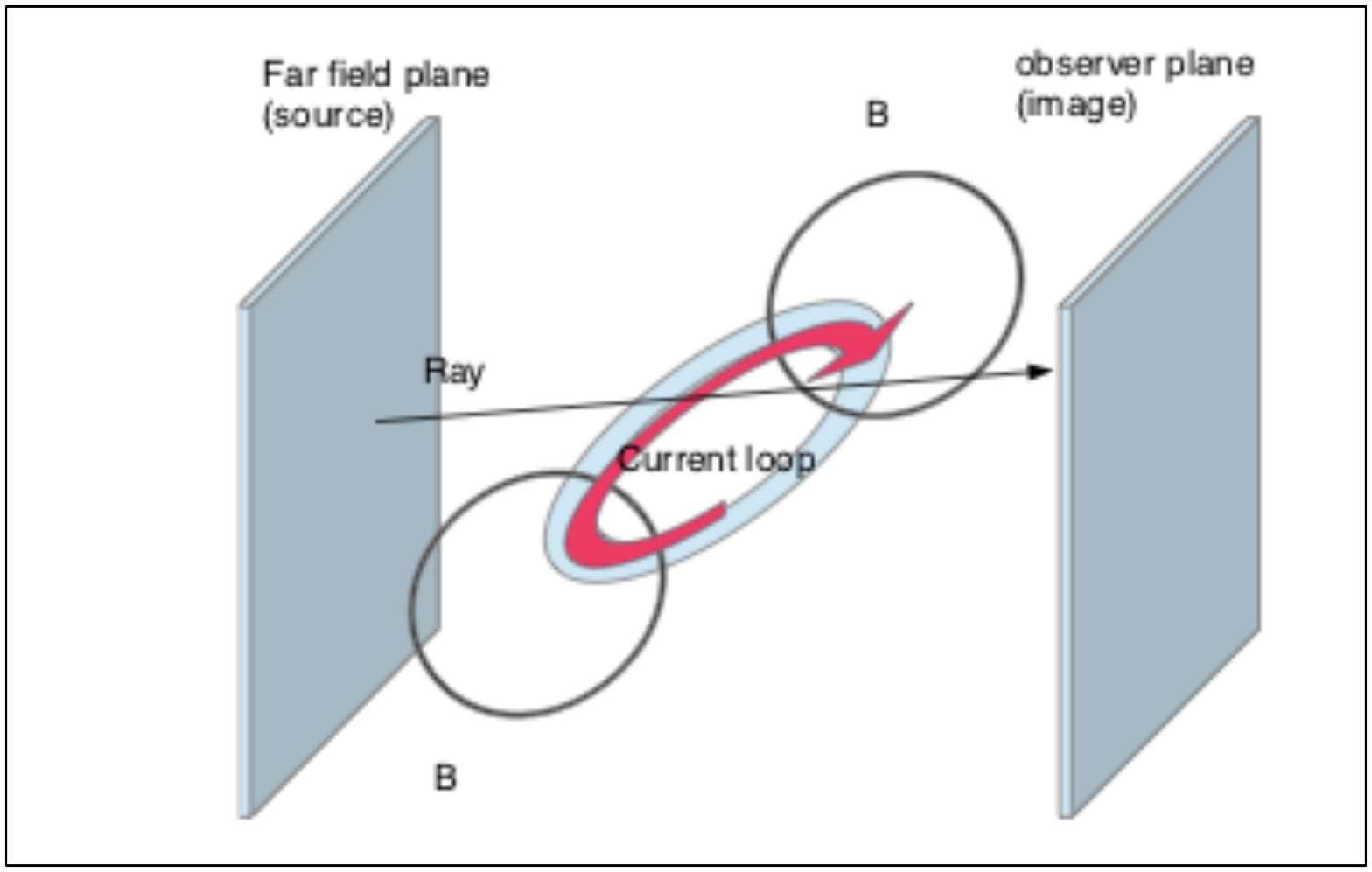}
\centering\includegraphics[width=1 \columnwidth]{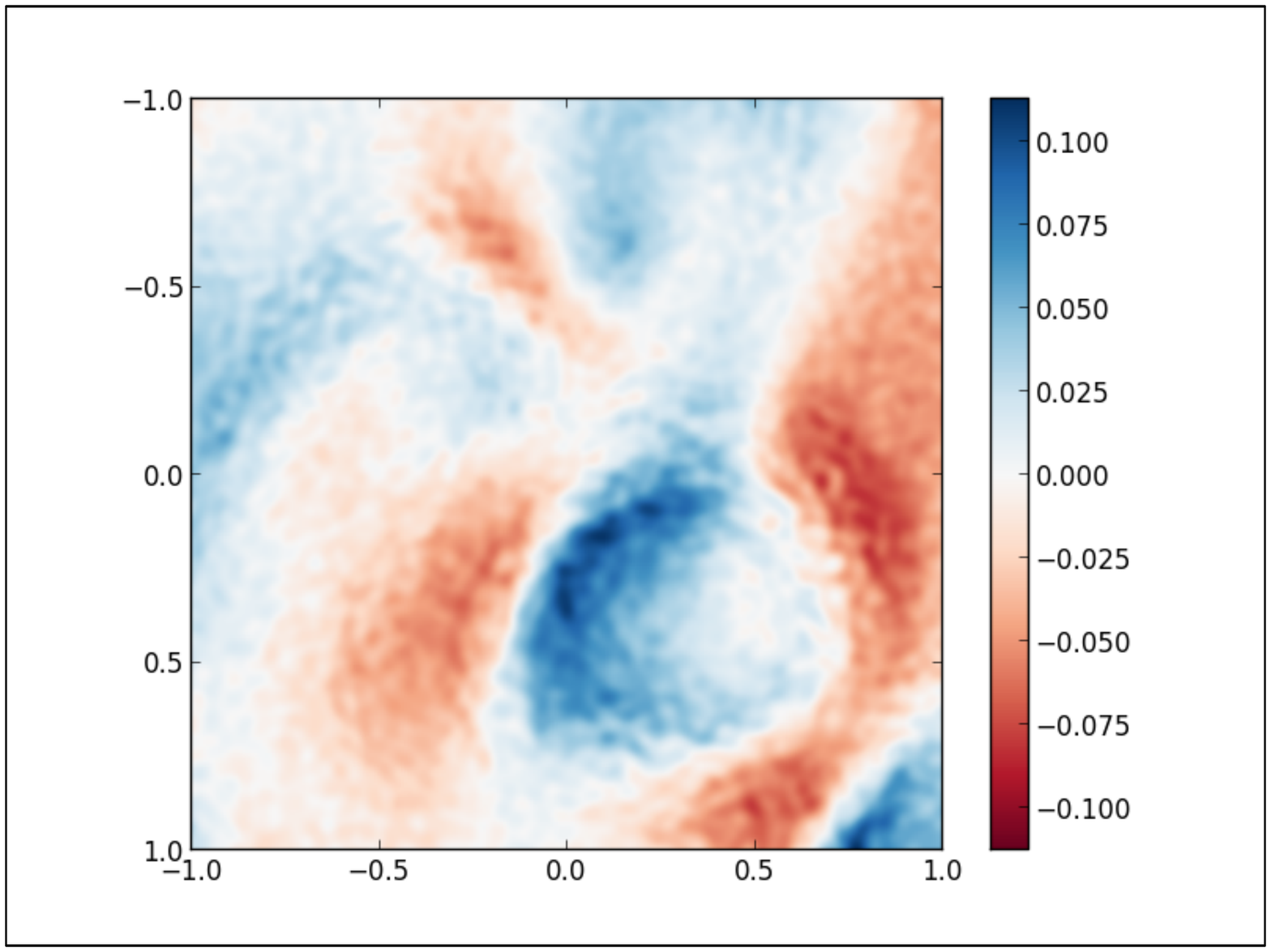}
\caption{Depolarization induced by a circular current loop, assuming a uniform initial ray distribution and that the dust density is uniform inside the torus: a) the geometry of the current loop; b) the final polarisation anisotropy observed far away.}
\label{fig2}
\end{figure}

\begin{figure}
\includegraphics[width=1 \columnwidth]{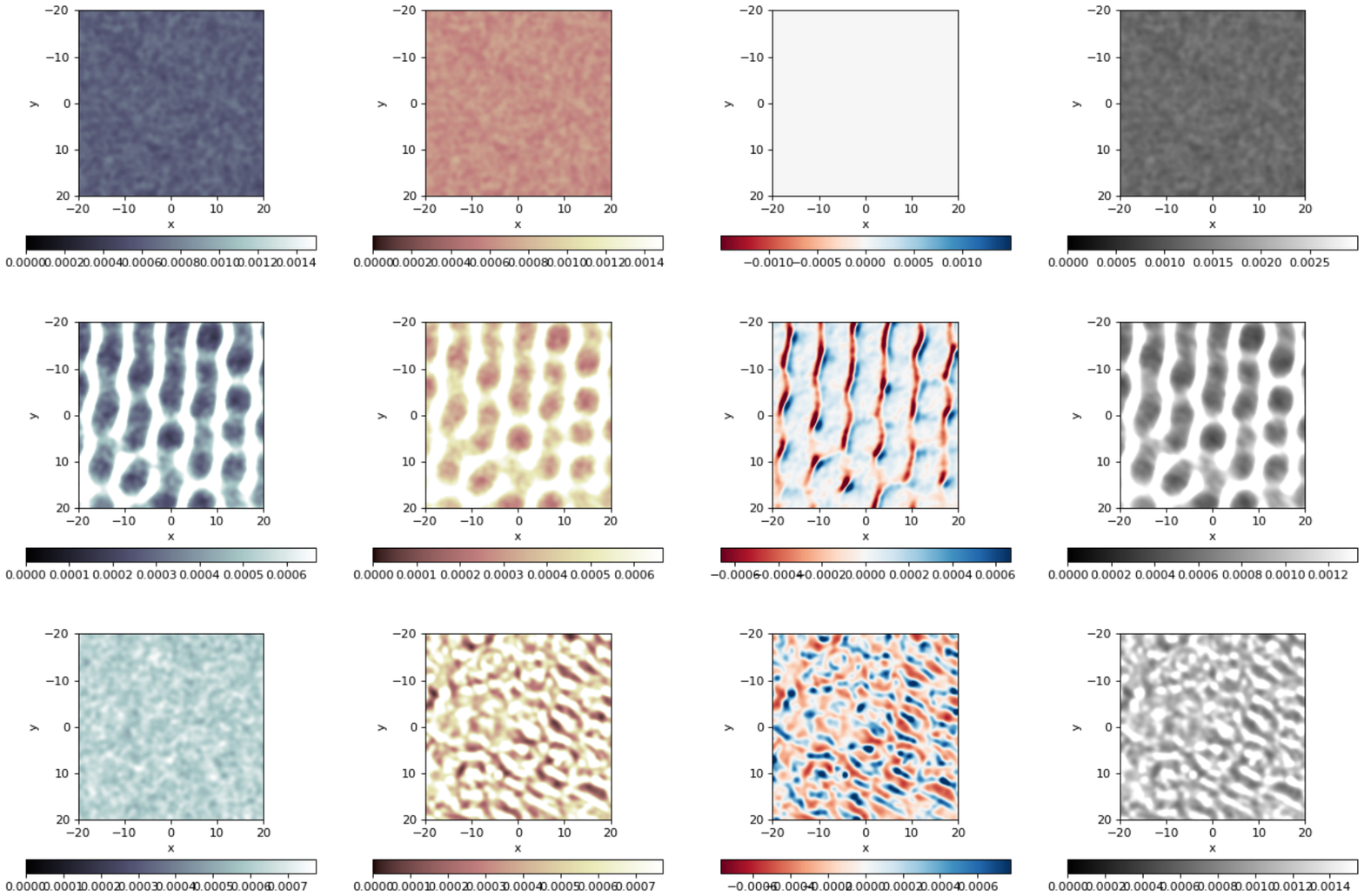}
\caption{Ray-tracing simulations of the CMB radiation propagating across the magnetized dust gas. The first column represents the intensity pattern of the ordinary mode $I_{ord}$, the second column corresponds to the intensity of the extraordinary mode, $I_{ext}$, the third column gives the difference between these two intensities $(I_{ord} - I_{ext}$, which approximately corresponds to the B-mode, and finally the fourth column indicates the total energy $(I_{ord} + I_{ext}$. The upper raw gives the initial conditions, which show some residual fluctuations inherent to the numerical method. The mid raw corresponds to large scale oscillations of the size of those observed in the CMB data, and the lower ray shows the results when some additional smaller scale oscillations are also added.}
\label{fig3}
\end{figure}

\begin{figure}
\includegraphics[width= 1 \columnwidth]{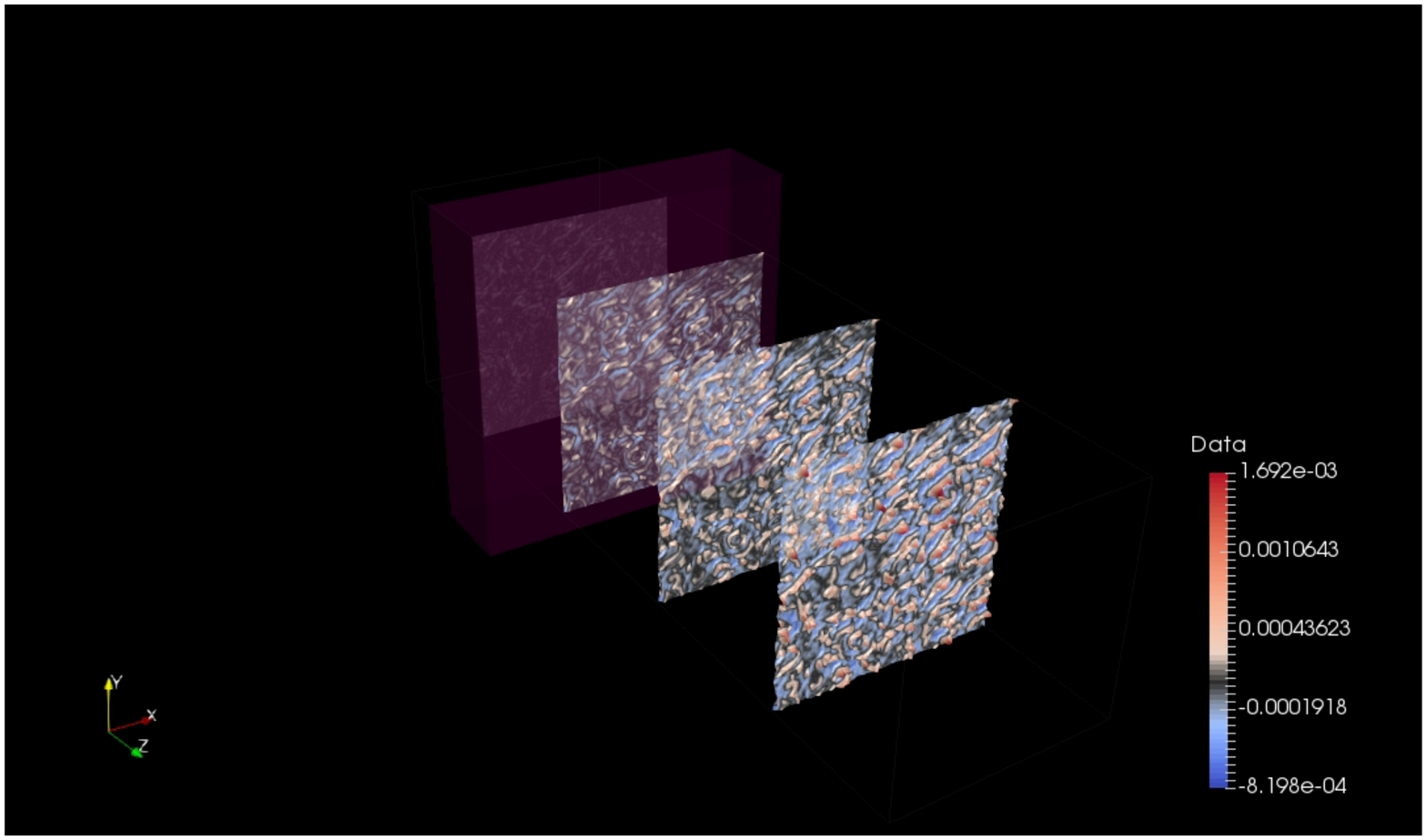}
\caption{Ray-tracing simulations of polarisation anisotropies created by density fluctuations of a magnetized dust gas: evolution of the B-mode pattern with the distance.}
\label{fig4}
\end{figure}

\section{Optical properties of the dust gas}

The equation of motion for the rotation of the grain dipole moment (represented with index q) in the presence of an electric field ${\bf E}({\bf r},t)$ is $(c=\mu_0 =\epsilon_0 =1)$.
\begin{equation}
\frac{d {\bf L}_q}{d t} = {\bf d}_q \times {\bf E}({\bf r}_q,t) - \gamma {\bf L}_q \, ,
\label{eq1} \end{equation}
where the last term on the right hand side describes an effective rotational drag with coefficient $\gamma$. We assume that there is a strong local magnetic field ${\bf B} = ( 0, 0, {\bf B}_0)$ that constrains the rotation of the dust to a plane, such that ${\bf d}_q = d (\cos \phi, \sin \phi, 0)$. The choice of this geometry can be understood in the following way \cite{shukla2002introduction, guerreiro2007dipolar}: the electric and magnetic moments are usually not aligned, and the magnetic moment is aligned along ${\bf B}$. Therefore, only the perpendicular component of the electric dipole 
moment can vary in time. Combining this result with the Maxwell equations
\begin{equation}
\nabla \times {\bf E}  = - \partial_t {\bf B} \, , \nabla \times {\bf H}  = \partial_t {\bf D} \, , \nabla \cdot {\bf D} = 0 \, ,  \nabla \cdot {\bf B} = 0  \, ,
\label{eq2} \end{equation}
the associated constitutive relations $({\bf D}  = \epsilon_0 \bar{\bar \epsilon}_{dust} {\bf E}$, and ${\bf B} =  \mu_0 {\bf H} )$, and taking an average over a small element of volume yields the following dielectric tensor 
\begin{eqnarray}
\bar{\bar {\epsilon}}_{dust} =  \left[
\begin{array}{ccc}
\epsilon^0_{dust} + \delta \epsilon d^2 /2  &  0 &  0 \\
 0 & \epsilon^0_{dust} + \delta \epsilon d^2 /2 &  0 \\
0  & 0  &   \epsilon^0_{dust} + \delta \epsilon d^2
\end{array}
\right]  .
\label{eq3} \end{eqnarray}
For a Fourier component of the electromagnetic field with frequency $\omega$, we can use
\begin{equation}
\epsilon^0_{dust} = 1 - \frac{\gamma^2 + 2 i \gamma \omega}{\omega^2}  , \quad
\delta \epsilon = \frac{\omega - i \gamma}{\omega^3} n_0 I_q^{- 1}  ,
\label{eq.4} \end{equation}
where $I_q$ is the dust moment of inertia.

The dielectric tensor $\bar{\bar{ \epsilon}}_{dust}$ has a degeneracy in terms of the principal axis, implying that the equivalent optical medium is uniaxial and the propagation of electromagnetic waves must be described in terms of ordinary and extraordinary waves, each with distinct polarisation states. The dielectric constants parallel and perpendicular to the principal axis are
\begin{equation}
\epsilon_{\parallel} = \epsilon^0_{dust} + \delta \epsilon d^2 , \quad  \epsilon_{\perp} = \epsilon^0_{dust} + \delta \epsilon d^2 / 2 .
\label{eq.5} \end{equation}
The principal axis of this effective medium is aligned with the magnetic field, which can typically vary over different regions of space. As a result, the fluid of dust exhibits local birrefrigence oriented by the local magnetic field (similar to the Cotton-Mouton effect) and characterised by the difference in parallel and perpendicular dielectric constants
$\Delta \epsilon = \epsilon_\parallel - \epsilon_\perp$. 
Since both $\epsilon^0_{dust}$ and $\delta \epsilon$ have real and imaginary parts, the refractive index and absorption coefficient is distinct for ordinary and extraordinary waves, and consequently $\Delta \epsilon$ must also have an imaginary part. As a result, the state of polarisation of light must necessarily change for a wave propagating in this medium, which can result in changes in the components of the B and E modes.

\section{Ray tracing}

In the simulations we apply the model developed by Sluijter et al. \cite{sluijter2008general}, where a ray is defined as the trajectory of the Poynting vector given by the integral curve of the Poynting vector field in terms of some parameter $\tau$ , as illustrated in Figure 1.
The trajectory of a beam of light propagating in a anisotropic medium can be obtained from the dispersion relation of the medium
$\omega ({\bf r}, {\bf p}) = \omega_{o(e)} ({\bf r}, {\bf p})$, with
\begin{eqnarray}
&\omega_o ({\bf r}, {\bf p}) = | {\bf p} |^2 - \epsilon_\perp \, , \\
& \omega_e ({\bf r}, {\bf p}) = \epsilon_\perp | {\bf p} |^2 - \epsilon_\perp (\epsilon_\perp + \Delta \epsilon)  + \Delta \epsilon ({\bf p} \cdot {\bf O} )^2 \, ,
\label{eq.6} \end{eqnarray}
being respectively the dispersion relations for the ordinary and extraordinary waves, and where ${\bf O}$ is a unit vector parallel to the local optical axis of the medium, usually refereed as the {\sl director}. Also, $ \epsilon_\perp = \epsilon_\perp ( {\bf r})$, $\Delta \epsilon = \Delta \epsilon ({\bf r})$ and ${\bf O} = {\bf O} ({\bf r})$ are assumed to be functions of the position, and ${\bf p}$ is the wavevector expressed in units of $c = 1$.
Then, the ray tracing equations can be written in canonical form as
\begin{equation}
\frac{d {\bf r}}{d \tau} = \alpha \frac{\partial \omega_{o/e}}{\partial {\bf p}} \, , \quad 
\frac{d {\bf p}}{d \tau} = - \alpha \frac{\partial \omega_{o/e}}{\partial {\bf r}} \, ,
\label{eq.7} \end{equation}
where $\alpha$ is an arbitrary function of order one, and influences only the parametric presentation of the ray position.
In this formulation, the ordinary and extraordinary electric polarization vectors can be obtained respectively as
\begin{equation}
{\bf e}_o = \frac{{\bf p} \times {\bf O}}{| {\bf p} \times {\bf O}|} \, , \quad 
{\bf e}_e = \frac{({\bf p} \times {\bf O}) \times \nabla_p \omega_e}{| ({\bf p} \times {\bf O}) \times \nabla_p \omega_e|} \, ,
\label{eq.8} \end{equation}
which allows us to compute the standard Stokes parameters, necessary to determined the components of the B and E modes. The number of rays used in each numerical simulation varies between 100 thousand and 400 thousand rays, with initial conditions chosen randomly in the perpendicular plan. The final image results from accumulation of rays on a grid of $256 \times 256$ cels. The resulting statistical error on each cell is of the order of 10 percent. 

The simulation box contains $300$ units, and the magnetic field and the dust grain density are assumed to follow a super-Gaussian distribution with a width corresponding to $20$ units. Such a distribution establishes a smooth boundary between the cloud and the empty region, and a uniform density is assumed inside the cloud. Density perturbations of up to $1$ percent are added to these super-Gaussian profiles. They are described by a random field with a Fourier spectrum obtained from an extrapolation of the observational data. 

We first explored the case where depolarisation is due to a magnetic field generated by a circular current loop.  In the simulations we  assume that  the intensity of the background radiation in the far field is randomly distributed, and on the average, is uniform. We also neglect radiation losses, which means that we neglect the eventual side scattering of both radiation modes by dust particles. Finally, the density of dust particles is a smoothed out torus, centred on the current loop. It should also be added that, in the simulations, we have not intended to match specific physical parameters, but only to illustrate the potential use of the model as a diagnostic tool for astrophysical phenomena. 

We assume the loop is at an angle of 45 degrees with respect to the direction pointing towards the observer, as shown in Figure 1a. The resulting depolarisation is represented in Figure 1b, where maxima and minima of polarisation anisotropy can be observed. This is similar to those observed in the so-called B-mode of the celebrated CMB measurements.

A second source of polarisation can be associated with the density granulation of the dust cloud, in a magnetic field background. In order to be more specific, we have made a space Fourier analysis of the CMB signal, and arrived at a two-dimensional pattern, which is represented in Figure 3, and the resulting granulation is assumed as a dust density or magnetic field granulation. Only static fluctuations are assumed, which means that eventual changes in the structure of the dust cloud can only occur in a time scale much larger than the escape time of radiation across the cloud. The resulting polarisation anisotropy is then calculated, and the results are shown in Figure 4. 

From left to right, this figure represents the intensity of the ordinary wave, the intensity of the extraordinary wave, the difference between these two intensities (the assumed B-mode) and the some of the two intensities (total intensity). The first row represents the initial conditions, assumed for the radiation spectrum, before entering the dust cloud. These initial conditions also show the level of noise associated with the ray tracing simulations. Only fluctuations well above this numerical noise is physically significant. In the mid row, we show the results obtained for a fluctuating dust density, in a fluctuating magnetic field, where the scales of both fluctuations are of the same order. Finally, in the last row, some short scale density fluctuations were added. Again, strong similarities with the observed B-mode of CMB radiation can be stated.

\section{Conclusions}

In conclusion, we have studied the polarisation changes induced by a magnetised dust cloud on a background spectrum of electromagnetic radiation. Such changes lead to an effect very similar to that of a gravitational delensing of the Cosmic Microwave Background, due to large gravitational structures in the universe. Formation of modulation structures similar to the B-mode can be identified in the CMB, after crossing an extended region of interstellar dust. This could be easily mistaken as a B-mode resulting from primordial gravitational waves.

This dust induced delensing results from a kind of collective Cotton-Mouton effect, which was described here with the help of a ray-tracing code for the two orthogonal polarization states associated with the magnetized dust gas. 
The ray-tracing simulations have also shown that, both magnetic field perturbations and dust density modulations can leave an imprint on the polarization pattern of the broad band radiation spectrum. For specific dust density modulation conditions, formation of caustics can take place, which however disappears with propagation. Our results could be useful to the understanding of CMB polarization. The importance of dust in the explanation of the observed B-mode signature was noted before \cite{adam2016planck}. Previous models were based on thermal radiation from dust grains. Here we provide an alternative and more specific process, which can be used to explain qualitatively the observed phenomena. This, in a sense, completes the demonstration that dust is the main ingredient leading to the formation of a B-mode on the polarization pattern of the CMB radiation. Specific models of dust grain distributions will have to be establish in order to bring the explanation to more a quantitative accuracy level.


\begin{acknowledgements}
We would like to thank Dr. Nuno Silva for his help with the preparation of figure 4.

\end{acknowledgements}

\bibliography{report}

\end{document}